\documentclass[journal]{IEEEtran}

\usepackage[font=scriptsize]{caption}
\usepackage{vcell}
\usepackage{rotating}
\usepackage{xcolor,soul,framed} 

\colorlet{shadecolor}{yellow}
\graphicspath{{../pdf/}{../jpeg/}}
\DeclareGraphicsExtensions{.pdf,.jpeg,.png}
\usepackage{colortbl}
\usepackage{multirow}
\usepackage{makecell}
\usepackage{hhline}
\usepackage{graphicx}
\usepackage[cmex10]{amsmath}
\usepackage{array}
\usepackage{mdwmath}
\usepackage{mdwtab}
\usepackage{eqparbox}
\usepackage{url}
\usepackage{amssymb}
\usepackage{caption}
\usepackage{subfigure}
\usepackage{siunitx}
\usepackage{gensymb}

\begin{document}
\bstctlcite{IEEEexample:BSTcontrol}
    \title{Imaging on the Edge: Mapping Object Corners and Edges with Stereo X-ray Tomography}
  \author{Zhenduo~Shang and
      Thomas~Blumensath

  \thanks{The authors are with the Institute of Sound and Vibration Research at the University of
Southampton, Southampton, United Kingdom.}} 

\markboth{
}{Shang and Blumensath : XXXXX}

\maketitle

\begin{abstract}
X-ray computed tomography is a powerful tool for volumetric imaging, where three-dimensional (3D) images are generated from a large number of individual X-ray projection images. Collecting the required number of low noise projection images is however time-consuming and so the technique is not currently applicable when spatial information needs to be collected with high temporal resolution, such as in the study of dynamic processes. In our previous work, inspired by stereo vision, we developed stereo X-ray imaging methods that operate with only two X-ray projection images. Previously we have shown how this allowed us to map point and line fiducial markers into 3D space at significantly faster temporal resolutions. In this paper, we make two further contributions. Firstly, instead of utilising internal fiducial markers, we demonstrate the applicability of the method to the 3D mapping of sharp object corners, a problem of interest in measuring the deformation of manufactured components under different loads. Furthermore, we demonstrate how the approach can be applied to real stereo X-ray data, even in settings where we do not have the annotated real training data that was required for the training of our previous Machine Learning approach. This is achieved by substituting the real data with a relatively simple synthetic training dataset designed to mimic key aspects of the real data.        

\end{abstract}

\begin{IEEEkeywords}
feature detection, X-ray Computed Tomography, stereo matching
\end{IEEEkeywords}

\IEEEpeerreviewmaketitle


\section{Introduction}


\IEEEPARstart{X}{R}ay Computed Tomography (XCT) is an established volumetric imaging technique used throughout medical, scientific and industrial applications. However, in order to generate volumetric images of high spatial resolution and with limited artefacts the method requires a very large number of individual X-ray measurements to be collected from around the object. Whilst the use of advanced algorithms, such as those based on machine learning or regularised optimisation (e.g using Total Variation (TV) constraints) offers the ability to reduce the number of required measurements somewhat  \cite{song2007sparseness, Lietal, yuan2018sipid, lee2018deep, pelt2022cycloidal, ernst2021sinogram, adler2018learned, li20213}, a significant reduction in the number of measurements is still not possible for most objects without sacrificing image quality.

We recently demonstrated a different approach \cite{shang2023stereo}. Instead of trying to reconstruct full tomographic images from limited observations, which requires very strong prior information, we instead recovered the 3D location of simple features, such as points and lines. We demonstrated that this could be done utilising only two (stereo) projection images, which was achieved without the strong constraints that are imposed in full image reconstruction from limited measurements. Our approach thus worked for general objects, as long as these include simple fiducial markers. This approach is of particular interest in time-sensitive applications, where the internal structure of an object changes rapidly, and where we might only be able to take a single pair of images at each time step. 

In this paper, we extend the above approach from fiducial markers to the corners and edges of the object, which is often more complex in terms of being able to identify point and line feature locations. There are several applications where this might be of interest, but where fiducial markers cannot be embedded into the object. Of particular interest to us are applications where we want to study a manufactured component with sharp corners and edges whilst it is undergoing rapid deformations.

As in our previous work, we assume an imaging setup with two or more X-ray sources and detectors, providing an X-ray stereo vision system. Using ideas from computer stereo vision, spatial mapping of object corners and edges then becomes possible at the speed of the detector frame rate, which is orders of magnitudes faster than full computed tomography data acquisition.

In contrast to visible light imaging, where the light registered at a specific location on a camera's image sensor is commonly associated with light reflecting from a single 3D point on the surface of an opaque object, X-ray intensity measured on an X-ray detector contain X-ray attenuation information from an entire line through the object \cite{kak2001principles}. Whilst the main challenge in traditional stereo vision lies in accurately aligning points between the two images forming a stereo pair \cite{horn1986robot}, for X-ray stereo imaging, not only does this matching step become more difficult, an additional challenge is the identification of the location of distinct 3D points in the 2D projected images \cite{shang2023stereo}.

\subsection{Our method}
Our previous work \cite{shang2023stereo} has identified an approach that firstly, identifies all point-like features in the two X-ray views, and secondly, matches these features between the views. Once identified and matched, mapping the features into 3D space then employed the same geometric considerations as traditional stereo vision. 
We use a block-based deep learning approach to identify the projected locations of 3D features in 2D space. These identified locations are then mapped into 3D space using the filtered back-projection (FBP) method \cite{sagara2010abdominal,hoffman1979quantitation}. 3D feature locations can then be identified as those points where individual back-projected features overlap between the two views in 3D space. Whilst this is unique with a high probability if the features are sparse and randomly distributed in space, for larger collections of features utilising additional views is an option. To further enhance the robustness against the exact localization of features on the two imaging planes, we here employ a second deep neural network to process the back-projected volumetric image to identify feature locations.

\subsection{Contributions}
Our previous approach is here extended to the mapping of corner and edge features. Whilst these features are conceptually similar to point and line fiducials, the difference is that they are more difficult to identify in X-ray projection images. Whilst using a fiducial marker with a higher X-ray attenuation value will produce projection images with discontinuous image intensities at the fiducial locations, this is not true for edges and corners, where the image intensity in the projected images changes smoothly at the feature locations. 
Therefore, detecting and matching in this case can pose greater difficulty. Not only do we demonstrate in this paper that our previous approach still works in this setting, we here make a second key contribution. In our previous paper, we used a machine learning method for feature identification that required training. In real applications, we do however seldom have sufficient amounts of real training data to train a model for a specific imaging task. To overcome this challenge, we here demonstrate that model training can also be done on simplified simulated data matched to a real imaging setting.

\section{Methodology}
 \begin{figure}[h!]
   \begin{center}
   \includegraphics[width=3.5in]{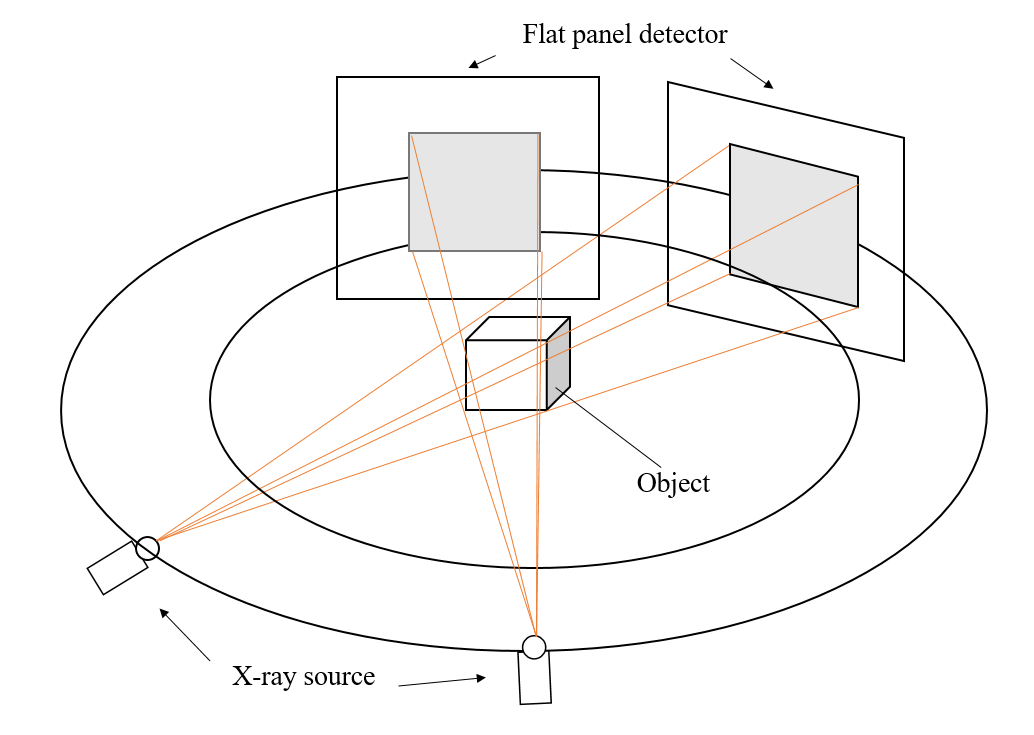}\\
   \caption{For stereo X-ray tomographic imaging with two views, two X-ray projection images are taken of an object from two different viewing directions.}\label{Two X-ray Sources Schematic}
   \end{center}
 \end{figure}
\begin{figure*}[t!]
  \begin{center}
  \includegraphics[width=6.5in]{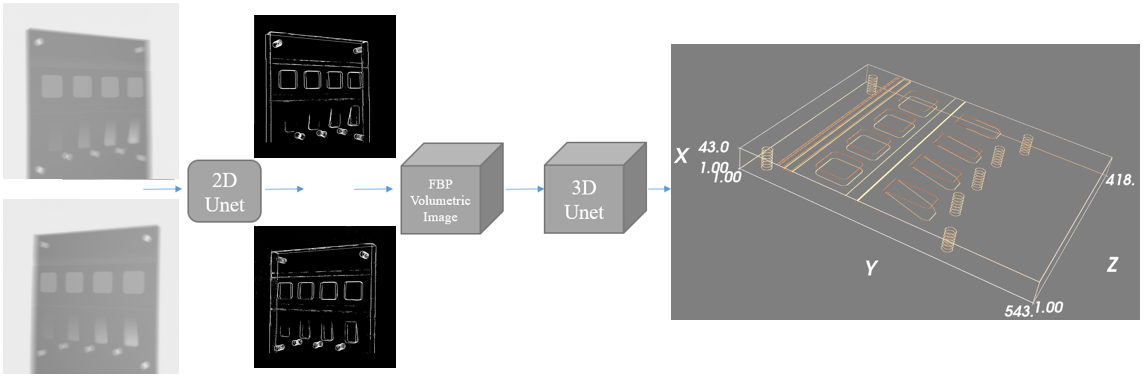}\\
  \caption{Overview of the proposed framework. The input is a pair of X-ray projection images. Each projection is fed independently into the same 2D U-net to compute two different feature maps, where the background is removed leaving estimates of the line and point feature locations. Utilising scan geometry knowledge, the two feature maps are then back-projected into a 3D volume using the FDK algorithm, with the back-projected volume further processed using a 3D U-net to generate the 3D spatial feature maps.}\label{overview of framework}
  \end{center}
\end{figure*}

The approach utilises a similar point identification and mapping process to our previous work \cite{shang2023stereo}. We assume a stereo X-ray tomography system as shown in \ref{Two X-ray Sources Schematic}. Features in each of the stereo images are identified, and mapped into 3D space, where the back-projected volume is used to identify 3D feature location. We summarise our proposed approach in Fig. \ref{overview of framework}.

\subsection{Feature Detection}

We formulate the feature detection problem as a binary classification problem. For a given projection image, we use a deep neural network that, for each pixel, estimates the probability that this pixel comes from a feature. A neural network implements a parameterised map  
 \begin{equation}
  \mathbf{y}_{mask}  =  f\left ( \mathbf{x}_{raw};\theta \right )\label{eq:1.1},
\end{equation}
where $\theta$ are the model parameters, $\mathbf{x}_{raw}\in \mathbb{R}^{m\times n}$ is the X-ray projection images encoding the spacial distribution of measured X-ray attenuation on one of the imaging planes and $\mathbf{y}_{mask}\in \mathbb{B}^{m\times n}$ is a pixel wise class probability map that can be thresholded to estimate feature locations. Parameters $\theta$ adapted using stochastic gradient optimisation to minimise the empirical risk over a training data-set $\left \{ \left ( \mathbf{x}^{i}_{raw},\mathbf{y}^{i}_{mask} \right ) \right \}^N_{i=1}$ which comprises $N$ image pairs. The function $f(\cdot)$ in Eq.\eqref{eq:1.1} is realised using the standard U-net architecture described in \cite{ronneberger2015u}, but implemented and trained as a classification network(i.e. using a sigmoid activation function and a binary cross-entropy loss).  

\subsection{Feature matching and 3D mapping}
As in \cite{shang2023stereo}, to derive a robust feature matching and 3D mapping approach, we use filtered back-projection methods to map the identified feature locations into 3D space. This is followed by a feature location identification step, where a deep neural network is applied to the 3D image to identify feature locations. Formally, if $B_L(\cdot)$ and  $B_R(\cdot)$ are the filtered back-projection operators \cite{feldkamp1984practical} for the left and right projection images\footnote{Note, extensions to settings with three or more projections follow the same ideas}, then we train a mapping $g(\cdot)$ that maps the two extracted feature maps $\widehat{\mathbf{y}}_{mask}^{L}$ and $\widehat{\mathbf{y}}_{mask}^{R}$  to a 3D tomographic volume $\mathbf{y}_{vol}\in \mathbb{R}^{m\times n \times o}$. 
\begin{equation}
   \mathbf{y}_{vol} = g\left (B_L( \widehat{\mathbf{y}}_{mask}^L)+ B_R(\widehat{\mathbf{y}}_{mask}^R);\theta  \right ) \label{eq:1.2}
\end{equation}
Here, $\widehat{\mathbf{y}}_{mask}^R$ and $\widehat{\mathbf{y}}_{mask}^L$ are the estimated 2D feature maps estimated from the left and right X-ray images and $\mathbf{y}_{vol}$ is the estimated volumetric image encoding the probability that each voxel contains one of the features. The function $g(\cdot)$ in Eq. \eqref{eq:1.2} is again parameterised by trainable parameters $\theta$. We here use the 3D U-net described in \cite{cciccek20163d} to implement this function.


\section{Dataset}

\begin{figure}[b!]
  \begin{center}
  \includegraphics[width=3.5in]{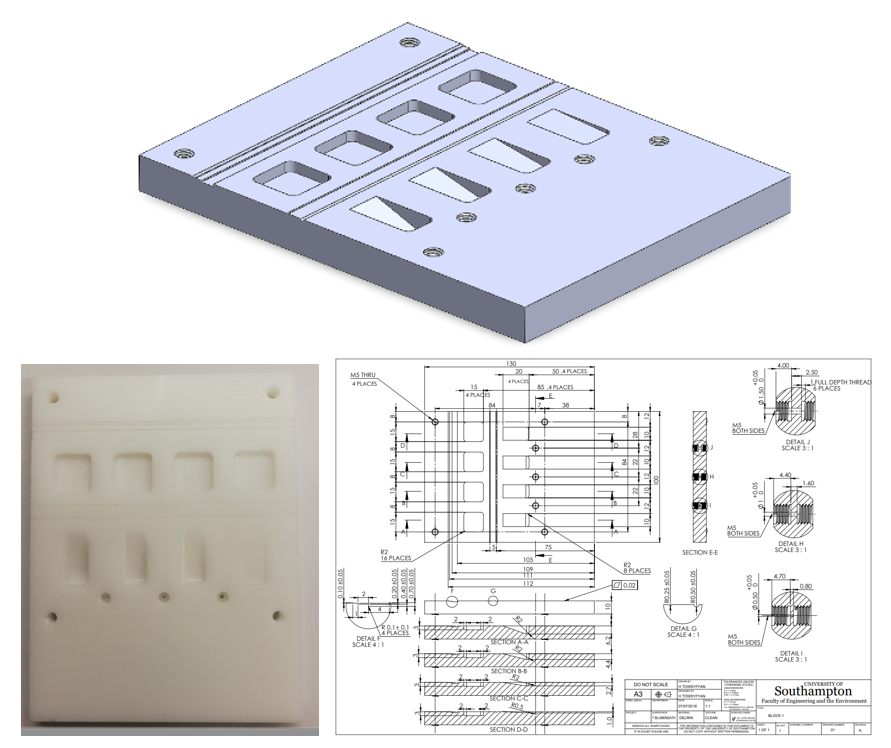}\\
  \caption{A Photo of the 3D block (left), 3D rendering (top) and its CAD drawing (right).}\label{3D block outlook}
  \end{center}
\end{figure}

We demonstrate the method's ability by mapping the edge and line features of a test phantom manufactured from homopolymer acetal. This phantom was imaged previously in unrelated work \cite{DEYHLE2020102222}. We show a photo of the object together with its original design drawing and a 3D rendering in Fig. \ref{3D block outlook}. 

\begin{figure}
  \centering
  \includegraphics[width=1.7in]{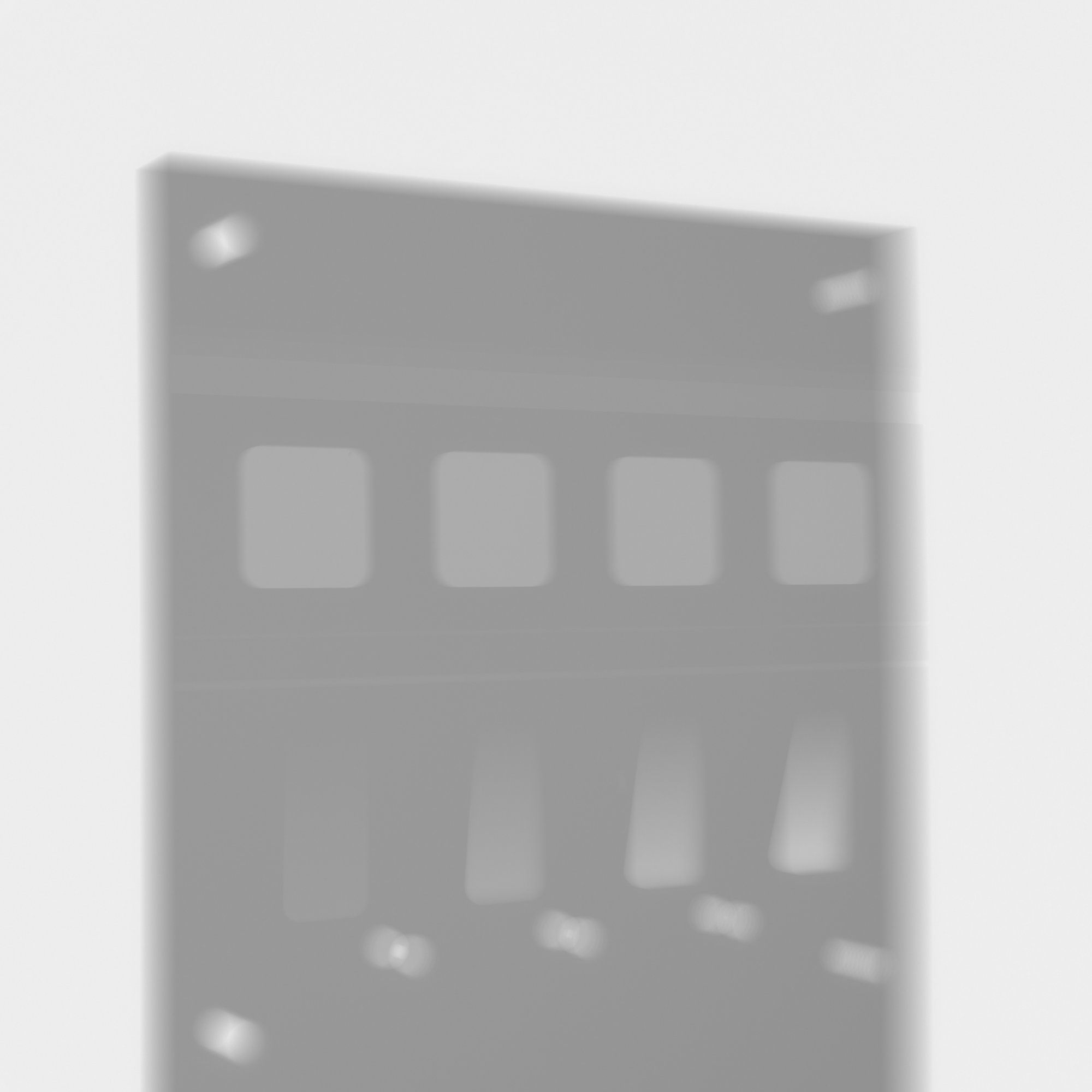}
  \includegraphics[width=1.7in]{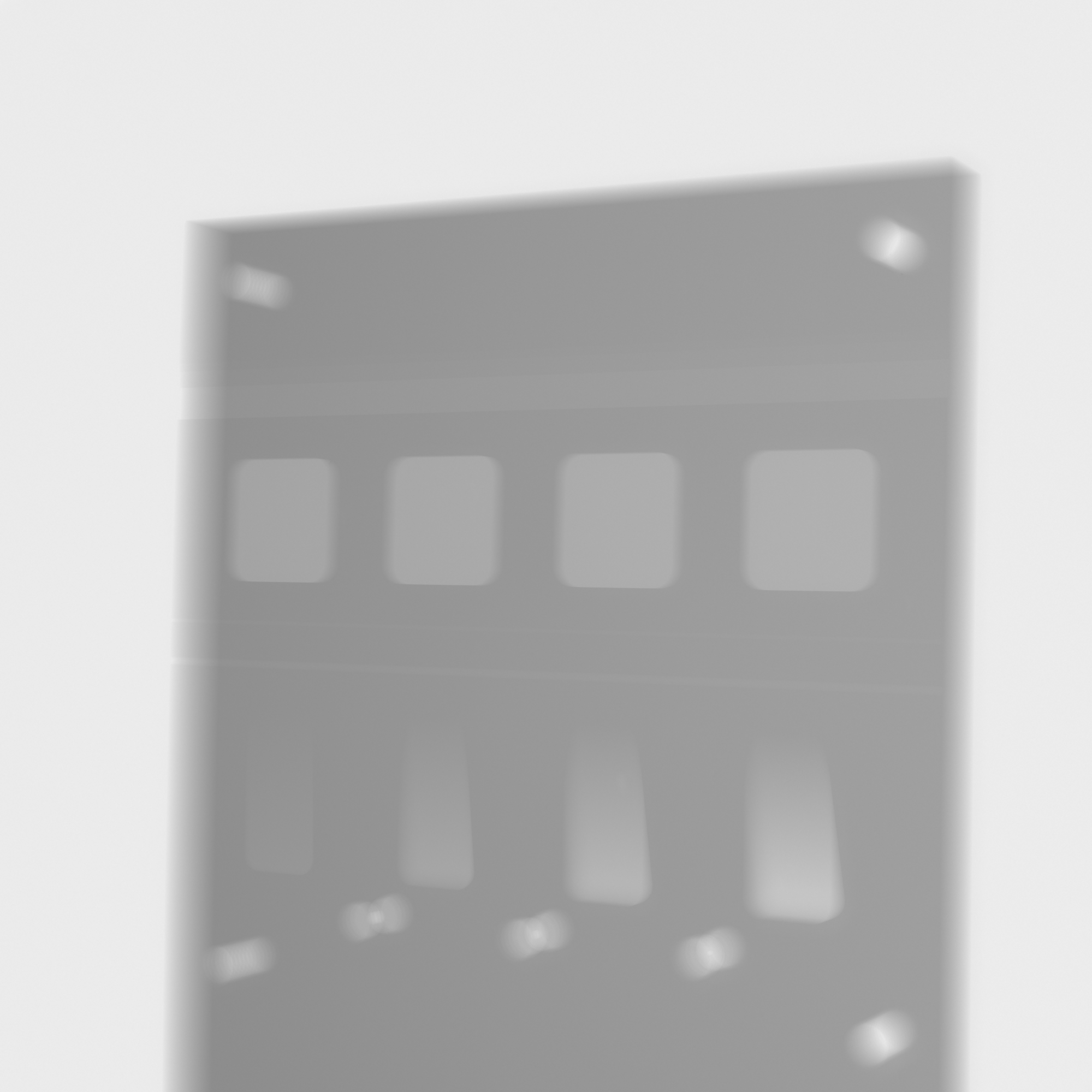}
  \caption{Intensity projection images of the physical test sample, collected by Nikon XTH225 X-ray tomography system with a $60\degree$relative rotation. }\label{feature detection test data}
\end{figure}

The object was originally scanned over a range of angles, though here we utilise only two projections taken at approximately $60\degree$. Scanning was performed on a Nikon XTH225 X-ray micro-tomography system. The original X-ray intensity images are shown in Fig \ref{feature detection test data}. The $2000\times 2000$ detector had a pixel size of 0.2mm and the object was scanned with a source-to-detector distance of about 923mm, and a source object distance of about 290mm.

\begin{figure}[b!]
  \begin{center}
  \includegraphics[width=3.5in]{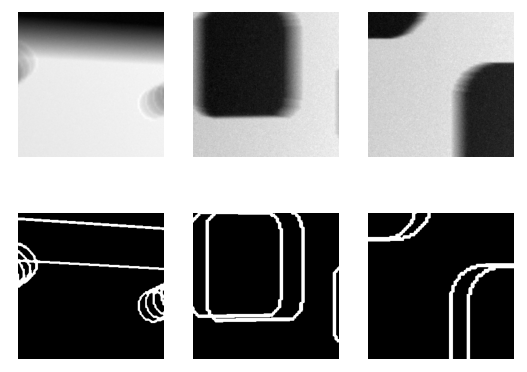}\\
  \caption{Three projections from the set of training samples (top) that were used to train the feature detection network. The top images show the simulated X-ray projection image blocks whilst the bottom images show the projected ground truth edge feature maps.}\label{training samples}
  \end{center}
\end{figure}

To train the feature detection 2D U-net model, we generated synthetic training datasets, generating 12 3D images of size $256\times256\times256$. Each image contained several simple 3D shapes with straight or rounded corners that approximately matched the shapes expected in the real X-ray images. We then positioned these shapes within a large rectangular prism, assigning low attenuation to the shapes and higher attenuation to the prism. Gaussian noise was added to the background. From these 12 3D images we generate 24 $1024\times1024$ 2D projection images at random object orientations using the Astra Toolbox \cite{van2015astra}. Each 2D projection was partitioned into 144 overlapping blocks of size $256\times256$, providing 3456 samples for training. We show three randomly selected 2D training data pairs in Fig. \ref{training samples}, where we show the projection images (top) together with the projected ground truth binary images identifying the locations of the corner and edge features (bottom).  

\begin{figure}
  \begin{center}
  \includegraphics[width=3.5in]{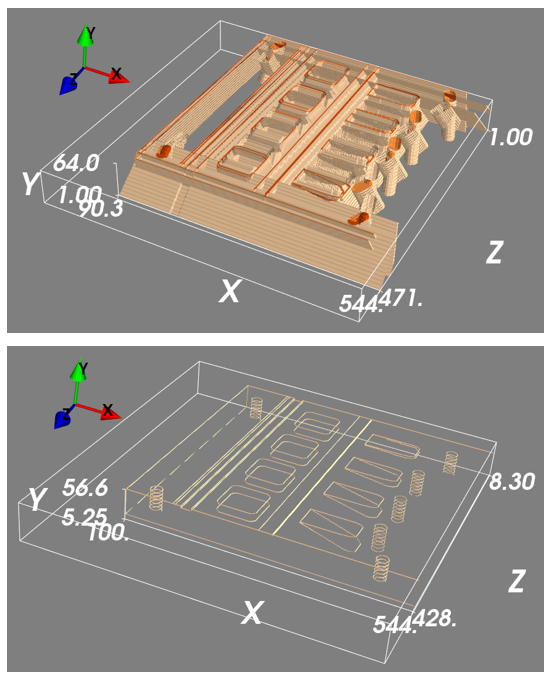}\\
  \caption{A set of 3D feature mapping training samples.The top is a thresholded 3D rendering of the back-projected volume that is used as the input to the machine learning model, and the bottom is the ground truth we are trying to predict.}\label{3D feature mapping training sample}
  \end{center}
\end{figure}

To train the 3D U-net to map the detected features in the projection images into 3D space, we generate a 3D edge map from its CAD drawing, which only contained object edge features. To generate a diverse set of images, the same edge feature map was rotated with $1\degree$ intervals around an axis parallel to the longest object side. This generated 360 3D images with edge features, each of size $544\times64\times544$ voxels and with a voxel size of $0.24\times0.24\times0.24 mm^3$. Each of these 3D blocks was then projected to generate pairs of 2D projection images, where projections were collected at $\pm 30\degree$. These ideal 2D feature maps were then back-projected into 3D images using the FDK algorithm to generate simulated back-projected feature maps. When training our 3D U-net we can thus use the simulated back-projected feature maps as network inputs, with the original clean edge feature maps as desired outputs. Example data is shown in Fig \ref{3D feature mapping training sample}.

\subsection{Calibration for stereo X-ray imaging system}
Whilst we had nominal values for the main system parameters, at the time of scanning, the system was not fully calibrated so that the nominal values given above might have significant errors. Thus, with only two raw real projections, by the epipolar constraint method \cite{ullman1979interpretation} with manually selected matching points from two real projections, the relative camera matrix between two cameras can be calculated. We here use the first view as the reference coordinate. Thus the first camera matrix can be denoted as Eq. \ref{eq:1.3}, where $P_{1}$ is the projection matrix of the first view, $K$ is the intrinsic matrix, \textbf{I} is the identity matrix. The second camera matrix is denoted as Eq. \ref{eq:1.4}, where $R$ is the relative rotation matrix and $t$ is the normalized translation matrix between two views.  

\begin{equation}
   P_{1}=K\left [ I | \textbf{0} \right ] \label{eq:1.3}
\end{equation}

\begin{equation}
   P_{2}=K\left [ R | t \right ] \label{eq:1.4}
\end{equation}

\begin{figure}
  \centering
  \includegraphics[width=1.7in]{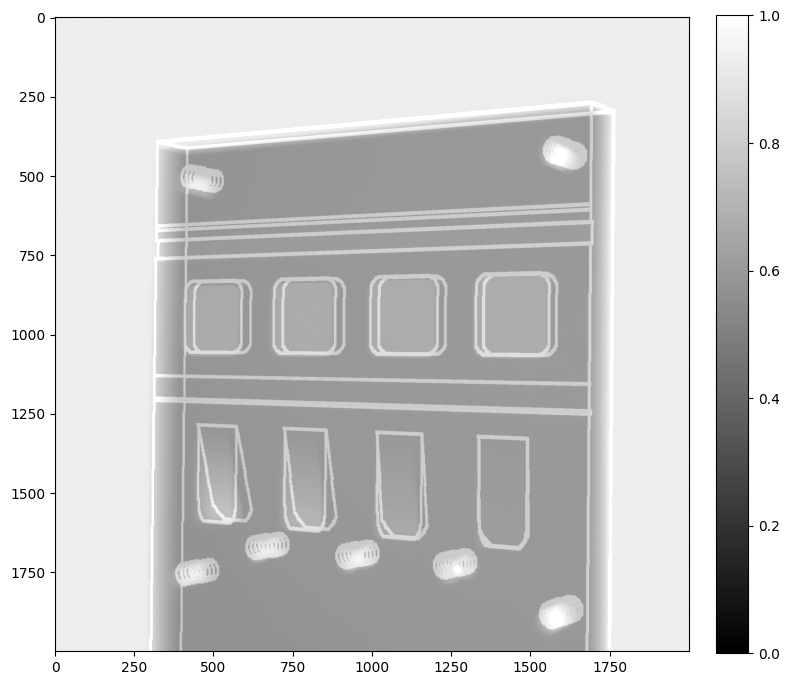}
  \includegraphics[width=1.7in]{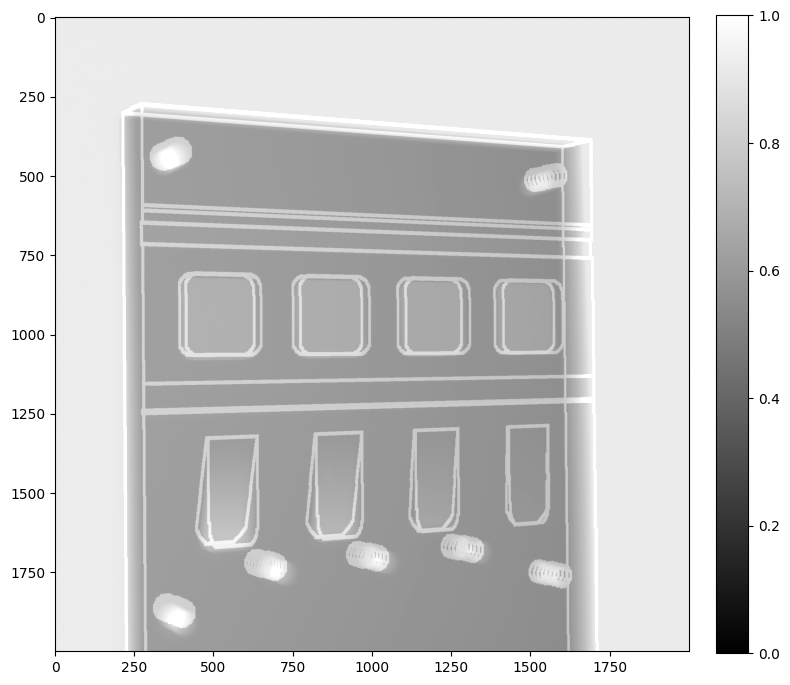}
  \caption{The overlap between simulated edge map projections and the real projections. We measured the geometric error at the object corners and the six screw holes positions, finding a 5 pixels error on average due to the geometric calibration procedure. }\label{simulated projections overlap on real}
\end{figure}

To further refine the calibration, the simulated 3D edge feature map was projected using the two estimated camera matrices and compared visually with the two real projections (see Fig \ref{simulated projections overlap on real}). We define the world coordinates based on the rotation centre of the stereo X-ray imaging system under the Astra Toolbox. Here we add a tiny pitch, roll and yaw to the centre point of the 3D block outline features to control its pose to make its forward projections under the stereo X-ray geometry mostly overlap with the real projections, while the two views angle are $-29\degree$ and $+32\degree$.The comparison of the real outline features and simulated projections is shown in Fig. \ref{simulated projections overlap on real}, demonstrating a good overlap of the simulated features and the real data after calibration.   

\section{Experimental evaluation}
Our stereo X-ray tomography approach is a two-step process, with feature detection and a 3D feature mapping step. Each of these will be evaluated here independently. 

\subsection{General training and evaluation approach}

Both networks were trained as classification networks for feature detection using the projections (2D network) or the filtered back-projection of the  2D projections (for the 3D network) as inputs and the binary images showing 2D projected or 3D point locations as output. Both networks were implemented using TensorFlow 2.x and optimised using an Nvidia GTX4070ti graphics card for 2D network and NVIDIA A100 graphics cards on the University of Southampton IridisX cluster for the 3D network. We use the Adam optimiser with the synthetic training data, a learning rate of $10^{-4}$ and 100 epochs. The loss function was the binary cross entropy.

\subsection{Feature Detection}

\begin{figure}
  \begin{center}
  \includegraphics[width=3.5in]{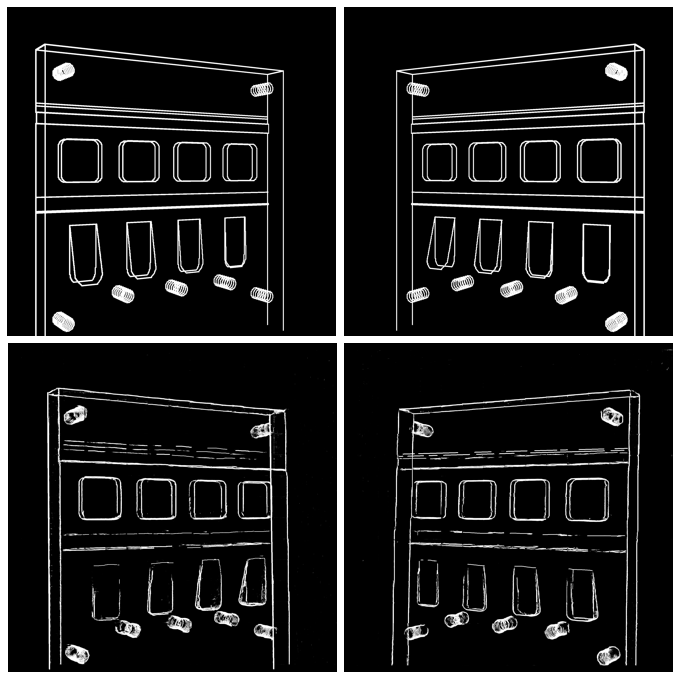}\\
  \caption{Comparison between the simulated ground truth 2D feature maps (top) and the 2D feature maps estimated from the real data (bottom).}\label{test performance}
  \end{center}
\end{figure}

After training the 2D feature detection U-net using the synthetic data described above as the training set, we used the real projection images from the physical phantom to test the method. The real test images were processed by converting the measured X-ray intensity into attenuation values before cropping the images into 338 overlapping blocks of size $256\times256$. We then applied the 2D U-net to all test sample blocks from both projections. Probabilities were averaged over all blocks that contained a particular image pixel before thresholding to produce a full size feature map. To evaluate the performance of the method, we use the simulated projections after calibration as our ground truth and visually compare the ground truth to the estimated 2D feature maps (See Fig. \ref{test performance}). 

\subsection{3D Feature mapping}
We then used the estimated 2D outline features of the real projections from the feature detection model and, for comparison, the two simulated feature projections in order to generate two back-projected volumes using the same calibrated projection geometry. Both back-projected volumes will go through the 3D model trained for 3D spatial position estimation.

\begin{figure}
  \begin{center}
  \includegraphics[width=3.5in]{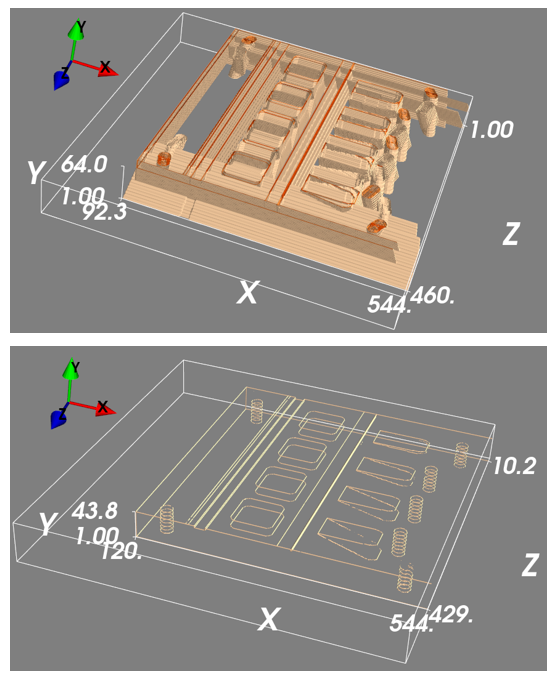}\\
  \caption{Visualisation of the operation of the 3D U-net. The top image is the back-projected volume generated from two simulated projections of the edge feature maps, whilst the bottom image is the 3D edge feature map estimated from the top image using the 3D U-net.}\label{3D mapping simulated result}
  \end{center}
\end{figure}

\begin{figure}
  \begin{center}
  \includegraphics[width=3.5in]{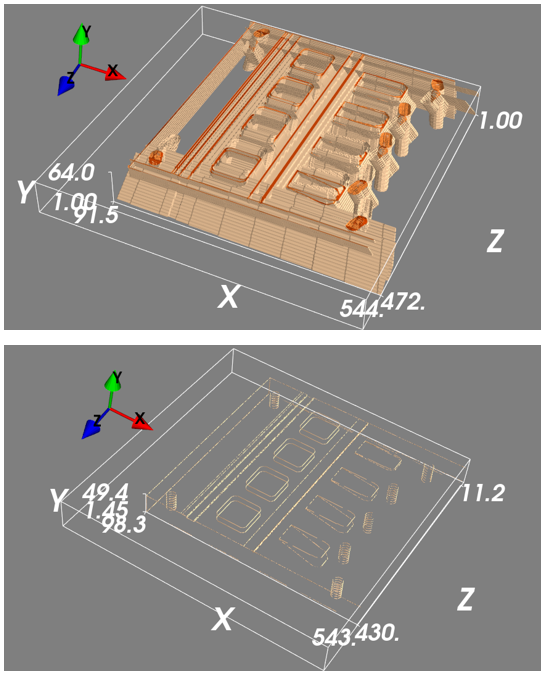}\\
  \caption{Back-projected feature maps and estimated 3D features for the real data.}\label{3D mapping real result}
  \end{center}
\end{figure}

\begin{figure}
  \begin{center}
  \includegraphics[width=3.5in]{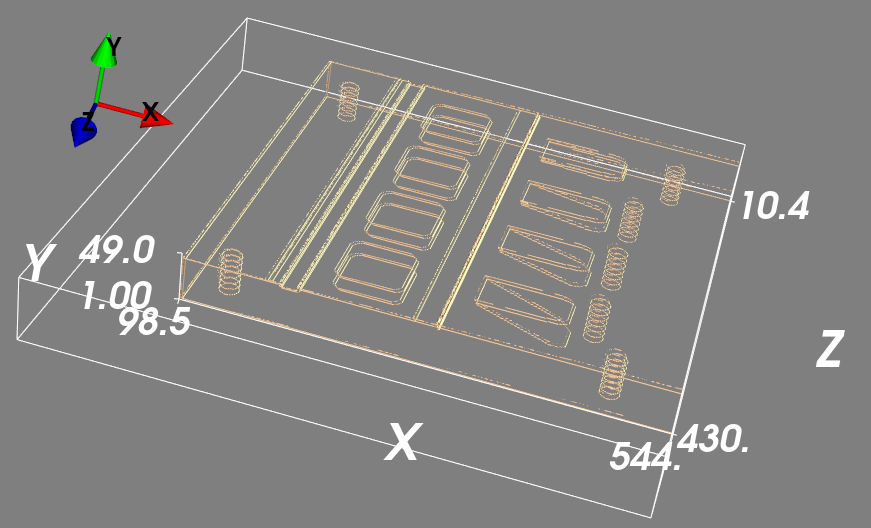}\\
  \caption{The positions error in 3D space between the 3D outline estimation by simulated projections and real outline features.}\label{err}
  \end{center}
\end{figure}

The back-projected volumes generated by the two simulated projections are shown in Fig. \ref{3D mapping simulated result} together with the 3D edge-feature map estimated with the 3D U-net model. This can be compared to the back-projected volume generated by the two 2D feature maps estimated from the real data shown in Fig. \ref{3D mapping real result}, where we also show the 3D edge-feature map estimated with the 3D U-net model. To numerically evaluate the error between these two estimations and the ground truth, we compare their corner and screw holes positions to those for the simulated phantom. As seen in Fig. \ref{err}, the errors from corners and screw holes are in the range of between 1 to 7 voxels, or $0.24mm$ to $1.68mm$. Considering the size of the 3D block and the inaccuracy from calibration, these are relatively small errors, which are assumed to be mainly due to the ad hoc post-scan calibration used here. Better calibration results could be obtained using a dedicated calibration object with a fixed stereo X-ray imaging setup. 

\section{Discussion and conclusions}
In this paper, we apply the stereo X-ray tomography framework from our previous work to the estimation of the 3D location of the corners of a 3D block. Using two deep neural networks, trained using simulated data, we could extract point and line feature locations from the two real projection images and, using the calibrated camera matrices of the system, project these back into 3D space. To identify feature location in 3D space, a further, 3D neural network was used again, trained on simulated data. The numerical evaluation of the results was based on the location of features in a nominal geometry that was also used for system calibration and should theoretically be aligned with the true object location. Distance errors in feature location of between $0.24mm$ to $1.68mm$ were found which were likely dominated by errors in the calibration process, though a detailed analysis of the dimensional accuracy of the approach will be left to a future study where a more controlled system calibration approach can be used.

\section{Acknowledgment} 
We would like to thank NVIDIA for the provision of an NVIDIA Titan GPU and the authors acknowledge the use of the IRIDIS High Performance Computing Facility, and associated support services at the University of Southampton, in the completion of this work.
\ifCLASSOPTIONcaptionsoff
  \newpage
\fi





\bibliographystyle{IEEEtran}
\bibliography{IEEEabrv,Bibliography}

\begin{thebibliography}{10}
\providecommand{\url}[1]{#1}
\csname url@rmstyle\endcsname
\providecommand{\newblock}{\relax}
\providecommand{\bibinfo}[2]{#2}
\providecommand\BIBentrySTDinterwordspacing{\spaceskip=0pt\relax}
\providecommand\BIBentryALTinterwordstretchfactor{4}
\providecommand\BIBentryALTinterwordspacing{\spaceskip=\fontdimen2\font plus
\BIBentryALTinterwordstretchfactor\fontdimen3\font minus \fontdimen4\font\relax}
\providecommand\BIBforeignlanguage[2]{{%
\expandafter\ifx\csname l@#1\endcsname\relax
\typeout{** WARNING: IEEEtran.bst: No hyphenation pattern has been}%
\typeout{** loaded for the language `#1'. Using the pattern for}%
\typeout{** the default language instead.}%
\else
\language=\csname l@#1\endcsname
\fi
#2}}
\renewcommand\BIBentryALTinterwordstretchfactor{4}

\bibitem{song2007sparseness}
J.~Song, Q.~H. Liu, G.~A. Johnson, and C.~T. Badea, ``Sparseness prior based iterative image reconstruction for retrospectively gated cardiac micro-ct,'' \emph{Medical physics}, vol.~34, no.~11, pp. 4476--4483, 2007.

\bibitem{Lietal}
\BIBentryALTinterwordspacing
H.~LI, S.~KAIRA, J.~MERTENS, N.~CHAWLA, and Y.~JIAO, ``Accurate stochastic reconstruction of heterogeneous microstructures by limited x-ray tomographic projections,'' \emph{Journal of Microscopy}, vol. 264, no.~3, pp. 339--350, 2016. [Online]. Available: \url{https://onlinelibrary.wiley.com/doi/abs/10.1111/jmi.12449}
\BIBentrySTDinterwordspacing

\bibitem{yuan2018sipid}
H.~Yuan, J.~Jia, and Z.~Zhu, ``Sipid: A deep learning framework for sinogram interpolation and image denoising in low-dose ct reconstruction,'' in \emph{2018 IEEE 15th International Symposium on Biomedical Imaging (ISBI 2018)}.\hskip 1em plus 0.5em minus 0.4em\relax IEEE, 2018, pp. 1521--1524.

\bibitem{lee2018deep}
H.~Lee, J.~Lee, H.~Kim, B.~Cho, and S.~Cho, ``Deep-neural-network-based sinogram synthesis for sparse-view ct image reconstruction,'' \emph{IEEE Transactions on Radiation and Plasma Medical Sciences}, vol.~3, no.~2, pp. 109--119, 2018.

\bibitem{pelt2022cycloidal}
D.~M. Pelt, O.~Roche~i Morg{\'o}, C.~Maughan~Jones, A.~Olivo, and C.~K. Hagen, ``Cycloidal ct with cnn-based sinogram completion and in-scan generation of training data,'' \emph{Scientific Reports}, vol.~12, no.~1, pp. 1--13, 2022.

\bibitem{ernst2021sinogram}
P.~Ernst, S.~Chatterjee, G.~Rose, O.~Speck, and A.~N{\"u}rnberger, ``Sinogram upsampling using primal-dual unet for undersampled ct and radial mri reconstruction,'' \emph{arXiv preprint arXiv:2112.13443}, 2021.

\bibitem{adler2018learned}
J.~Adler and O.~{\"O}ktem, ``Learned primal-dual reconstruction,'' \emph{IEEE transactions on medical imaging}, vol.~37, no.~6, pp. 1322--1332, 2018.

\bibitem{li20213}
X.~Li, S.~Wang, P.~Chen, and L.~Wang, ``3-d inspection method for industrial product assembly based on single x-ray projections,'' \emph{IEEE Transactions on Instrumentation and Measurement}, vol.~70, pp. 1--14, 2021.

\bibitem{shang2023stereo}
Z.~Shang and T.~Blumensath, ``Stereo x-ray tomography,'' \emph{IEEE Transactions on Nuclear Science}, 2023.

\bibitem{kak2001principles}
A.~C. Kak and M.~Slaney, \emph{Principles of computerized tomographic imaging}.\hskip 1em plus 0.5em minus 0.4em\relax SIAM, 2001.

\bibitem{horn1986robot}
B.~Horn, B.~Klaus, and P.~Horn, \emph{Robot vision}.\hskip 1em plus 0.5em minus 0.4em\relax MIT press, 1986.

\bibitem{sagara2010abdominal}
Y.~Sagara, A.~K. Hara, W.~Pavlicek, A.~C. Silva, R.~G. Paden, and Q.~Wu, ``Abdominal ct: comparison of low-dose ct with adaptive statistical iterative reconstruction and routine-dose ct with filtered back projection in 53 patients,'' \emph{American Journal of Roentgenology}, vol. 195, no.~3, pp. 713--719, 2010.

\bibitem{hoffman1979quantitation}
E.~J. Hoffman, S.-C. Huang, and M.~E. Phelps, ``Quantitation in positron emission computed tomography: 1. effect of object size.'' \emph{Journal of computer assisted tomography}, vol.~3, no.~3, pp. 299--308, 1979.

\bibitem{ronneberger2015u}
O.~Ronneberger, P.~Fischer, and T.~Brox, ``U-net: Convolutional networks for biomedical image segmentation,'' in \emph{International Conference on Medical image computing and computer-assisted intervention}.\hskip 1em plus 0.5em minus 0.4em\relax Springer, 2015, pp. 234--241.

\bibitem{feldkamp1984practical}
L.~A. Feldkamp, L.~C. Davis, and J.~W. Kress, ``Practical cone-beam algorithm,'' \emph{Josa a}, vol.~1, no.~6, pp. 612--619, 1984.

\bibitem{cciccek20163d}
{\"O}.~{\c{C}}i{\c{c}}ek, A.~Abdulkadir, S.~S. Lienkamp, T.~Brox, and O.~Ronneberger, ``3d u-net: learning dense volumetric segmentation from sparse annotation,'' in \emph{International conference on medical image computing and computer-assisted intervention}.\hskip 1em plus 0.5em minus 0.4em\relax Springer, 2016, pp. 424--432.

\bibitem{DEYHLE2020102222}
\BIBentryALTinterwordspacing
H.~Deyhle, H.~Towsyfyan, A.~Biguri, M.~Mavrogordato, R.~Boardman, and T.~Blumensath, ``Spatial resolution of a laboratory based x-ray cone-beam laminography scanning system for various trajectories,'' \emph{NDT \& E International}, vol. 111, p. 102222, 2020. [Online]. Available: \url{https://www.sciencedirect.com/science/article/pii/S096386951930324X}
\BIBentrySTDinterwordspacing

\bibitem{van2015astra}
W.~Van~Aarle, W.~J. Palenstijn, J.~De~Beenhouwer, T.~Altantzis, S.~Bals, K.~J. Batenburg, and J.~Sijbers, ``The astra toolbox: A platform for advanced algorithm development in electron tomography,'' \emph{Ultramicroscopy}, vol. 157, pp. 35--47, 2015.

\bibitem{ullman1979interpretation}
S.~Ullman, ``The interpretation of structure from motion,'' \emph{Proceedings of the Royal Society of London. Series B. Biological Sciences}, vol. 203, no. 1153, pp. 405--426, 1979.

\end{thebibliography}

\vfill


\end{document}